\documentclass[12pt]{article}

\usepackage{latexsym}
\usepackage{graphics}
\newcommand{\be}{\begin{equation}}
\newcommand{\ee}{\end{equation}}
\newcommand{\ben}{\begin{eqnarray}}
\newcommand{\een}{\end{eqnarray}}
\newcommand{\bea}{\begin{eqnarray}}
\newcommand{\eea}{\end{eqnarray}}

\begin{document}

\setlength{\baselineskip}{19pt}
\title{
\normalsize \mbox{ }\hspace{\fill} \\[1ex]
{\large\bf Hawking radiation and Quasinormal modes
 \\[1ex]}}
\author{SangChul Yoon
\\
\\
{\it Physics Department, 104 Davey, Penn State,} \\
{\it University Park, PA 16802, USA}\\
}

\maketitle

\thispagestyle{empty}

\begin{abstract}
The spectrum of Hawking radiation by quantum fields in the curved
spacetime is continuous, so the explanation of Hawking radiation
using quasinormal modes can be suspected to be impossible. We find
that quasinormal modes do not explain the relation between the state
observed in a region far away from a black hole and the short
distance behavior of the state on the horizon.
\end{abstract}

Loop quantum gravity \cite{Rovelli et all} is a background
independent, non-perturbative approach to unify general relativity
and quantum physics. There has been a lot of progress, especially in
the resolution of the big-bang singularity \cite{Martin}, even
though we have physical states only in the kinematical level in loop
quantum gravity and solving the hamiltonian constraint
\cite{Thiemann1} is still a open problem\footnote{Even though it
still remains if physical states constructed with the master
constraint operator contain correct semiclassical states, Thiemann
showed the existence of the self-adjoint, positive master constraint
operator for loop quantum gravity and we have a good chance to solve
this open problem \cite{Thiemann2}. See also \cite{Muxin}.} and the
description of low energy physics \cite{Yoon} is not fully
understood yet.

In loop quantum gravity, there is the Immirzi parameter $\gamma$
\cite{Immirzi}. Neither phase space variables nor their Poisson
brackets depend on this parameter. Thus the canonical phase space is
$\gamma$ independent\footnote{This parameter does not appear in the
equations of motion. Recently it was shown that in the presence of
fermions, it appears in the equations of motion \cite{Perez}.}.
Therefore there is no ambiguity in quantization \cite{Ashtekar}.
However the expression of the geometrical fields -the spatial triad
and the extrinsic curvature- in terms of canonical variables depends
on it, so to fix the value of $\gamma$ is important to figure out
the correct semiclassical limit.

One of the ways to fix the value of the Immirzi parameter was
proposed using SO(3) gauge group instead of SU(2) \cite{Dreyer}.
This was motivated by observation \cite{Hod} on the quasinormal
modes of a black hole with using Bohr's correspondence principle:
``Transition frequencies at large quantum numbers should be equal to
classical oscillation frequencies." Even though a fair argument was
made to specify the imaginary part of quasinormal mode as a quantum
number in term of the relaxation time \cite{Hod}, there is a
criticism for using  the correspondence principle
\cite{Khriplovich}. If we can find any quantum mechanical role of
the quasinormal modes, above observation would be supported without
the correspondence principle. However we find that there is no
Hawking radiation by quasinormal modes in the framework of quantum
field theory in curved spacetime.

Fredenhagen and Haag derived Hawking radiation \cite{Hawking1} by
the local behavior of the correlation functions, $< \phi(x_1)\cdot
\cdot \cdot \phi(x_n) >$ , of the quantum field \cite{Fredenhagen}.
What they found was that the asymptotic counting rate of the
two-point correlation function of the smeared quantum field living
on a Schwarzschild black hole background is governed by the short
distance behavior of the ground state two-point function near the
horizon, so-called Hadamard form \cite{Wald}. The smeared field

\bea   Q=\int \phi(x)h(x)\sqrt{|g|}d^4x \eea where the test function
$h$ is a smooth function with support in the region far away from
the horizon. Expression of the counting rate $<Q^*_TQ_T>$ involves
the function $f^T$ as

\bea
          f^T(t,x)=\int dt_0f^{T,t_0}(t,x)
          \eea
where $T$ is used as the index of a sequence $Q_T$ which are time
translates with respect to the Schwarzschild time $t$ and
$f^{T,t_0}$ is the solution of the wave equation with the initial
conditions

\bea
           f^{T,t_0}(T+t_0,x)=0,&&\frac {\partial}{\partial
           t}f^{T,t_0}(T+t_0,x)=h(t_0,x)
 \eea

With this

\bea <Q^*_T Q_T>=\int\limits_{\tau_1=\tau_2=0}W^{(2)}(x_1,x_2) (D_1
\leftrightarrow) (D_2 \leftrightarrow)
f^T(x_1)f^T(x_2)dr_1dr_2r^2_1d\Omega_1 r^2_2d\Omega_2 \eea where
$\tau=t+r^*-r$, $r^*$ is the Regge-Wheeler tortoise coordinate
\cite{Wheeler}, and

\bea
D_i=\Big(1+\frac{r_0}{r}\Big)\frac{\partial}{\partial\tau_i}-\frac{r_0}{r}\frac{\partial}{\partial
r_i} \eea

\bea
W^{(2)}(x_1,x_2)=(2\pi)^{-2}\sigma^{-1}_{\varepsilon}+\hat{w}^{(2)}
\eea where $r_0$ is the Schwarzschild radius, $\sigma_{\varepsilon}$
is the square of the geodesic distance between $x_1$ and $x_2$, and
$\hat{w}^{(2)}$ is the less singular part of $W^{(2)}(x_1,x_2)$
which has the Hadamard form.
 After separating off the angular part, $f^T$ can be obtained by
inverse Laplace transformation

\bea
    f^T(t,r^*)=(2\pi i )^{-1} \int^{c+i \infty}_{c-i \infty} dz
    \tilde{f}(z,r^*)e^{-z(t-T)}, &&c>0
\eea where $t<T+t_1$ with $t_1=inf\{t_0|(t_0,r^*)  \in$  support of
$h\}$ and $\tilde{f}$ is the solution of the ordinary inhomogeneous
differential equation by Laplace transformation of the original
solution of the wave equation

\bea ( \frac{d^2}{dr^{*2}}-z^2-V_l)\tilde{f}(z,r^*)=F(z,r^*)=
    \int
dt_0 e^{zt_0}h(t_0, r^*) \eea with
 \bea V_l(r)=\Big(1-\frac{r_0}{r}\Big)\Big(\frac{l(l+1)}{r^2}+\frac{r_0}{r^3}\Big)
 \eea where $l$ is a index of
 angular momentum.

What is important in description of Hawking radiation is the
behavior of $f^T$. This  can be decomposed into three parts

\bea f^T = f^T_{-} + f^T_{+} + \triangle^T \eea As $T-t \rightarrow
\infty$, $f^T_{-}$ is accumulating on the horizon and $f^T_{+}$ is
moving to spatial infinity and the third term vanishes. The counting
rate is dominated by $f^T_{-}$ and is governed by the short distance
behavior of the state near the horizon \cite{Fredenhagen}.

We can express the counting rate with quasinormal modes. We can
write eq.(7) with quasinormal modes by closing the contour to the
left plane

\bea f^T(t,r^*) \sim \sum
 \limits_{q} Residue \left [
\tilde{f}(z_q,r^*) \right ] e^{-z_q(t-T)}   \eea where $z_q$ are
quasinormal mode frequencies\footnote{There are other contributions
by the branch cut from the singularity at $z=0$, which dominates at
late time and the left quarter-circle term which is not important
because it explains the signal at early time \cite{Nollert}.}. The
solution of eq.(8) can be obtained by the Green function method. Let
$G_+$, $G_{-}$ be the solutions of the homogeneous part of eq.(8)
with respective boundary conditions \bea G_{-}\rightarrow e^{zr^*},
r^*\rightarrow - \infty;&G_+ \rightarrow e^{-zr*}, r^*\rightarrow +
\infty  \eea In term of $G_+$ and $G_{-}$, $\tilde{f}$ is given by
\bea \tilde{f}(z,r^*)=\frac{1}{\delta(z)}\Big \{ G_{+}(z,r^*)
\int^{r^*}_{ - \infty}dr^* G_{-}(z,r^*)F(z,r^*)\nonumber
\\+G_{-}(z,r^*) \int^{\infty}_{r^*}dr^* G_{+}(z,r^*)F(z,r^*)\Big \}
\eea where $\delta(z)$ is the Wronskian between $G_{-}$ and $G_{+}$,
\bea \delta(z)= G_{+}\frac{\partial G_{-}}{\partial
r^*}-\frac{\partial G_{+}}{\partial r^*} G_{-}.\eea Quasinormal mode
frequencies $\{z_q\}$ are given by zeros of the Wronskian
$\delta(z)$ and at $z=z_q$, $G_{-}$ and $G_{+}$ are equal up to a
constant factor.

To express $<Q^*_T Q_T>$ as the sum of each quasinormal mode, we
insert eq.(7) to eq.(4), interchange the order of the integral in
eq.(7) and the integral in eq.(4) and closing the contour of eq.(7).
We can not insert eq.(11) to eq.(4) and interchange the order of the
sum in eq.(11) and the integral in eq.(4) because each mode behaves
as $e^{2z_q r^*}$ around the horizon and the integral diverges.
Therefore without regularization, we can not convince that $<Q^*_T
Q_T>$ as the sum of each quasinormal mode converges. If we take the
interval of $r$ in eq.(4) between $r
> r_0$ and $\infty$, and take $r$ goes $r_0$ after taking $T$ goes
$\infty$, we can see that $<Q^*_T Q_T>=0$. In general, if $<Q^*_T
Q_T>$ converges, it is 0 by the exponential decaying factor of
$e^{z_q T}$.

We saw that Hawking radiation at late time is not explained by
quasinormal modes. At finite time, eq.(11) is a very good
approximation, but $f^T$ is not dominated by the asymptotic
frequency which decays very fast. In addition to this, we do not see
any special role of the horizon. Even though the counting rate can
be written by the quasinormal modes expansion, we can not interpret
this as particle creation by black hole.

The author is grateful to  Abhay Ashtekar and Jonathan Engle for
collaboration and useful discussions at the early stage of this work
\footnote{We tried to derive Hawking temperature with quasinormal
modes, which was not successful. This paper shows that it is
impossible to derive Hawking temperature with quasinormal modes.}.


\end{document}